\documentclass[conference,final]{IEEEtran}

\usepackage{epsfig}     
\usepackage{url}        
\usepackage{amsmath}
\usepackage{multirow}
\usepackage{subfig}

\newcommand{\refFig}[1]{Figure~\ref{#1}}
\newcommand{\refTab}[1]{Table~\ref{#1}}
\newcommand{\refSec}[1]{Section~\ref{#1}}
\newcommand{\etal}{\emph{et~al.}}
\newcommand{\scap}[1]{\emph{\small{#1}}}

\IEEEoverridecommandlockouts
\begin{document}

\title{Can Realistic BitTorrent Experiments Be Performed on Clusters?}
\author{
\IEEEauthorblockN{Ashwin Rao$^{\dagger}$, Arnaud Legout, and Walid Dabbous}
\IEEEauthorblockA{INRIA, France.\\
\{ashwin.rao, arnaud.legout, walid.dabbous\}@inria.fr}
\thanks{$^{\dagger}$This is the author version of the paper published in the Proceedings of the $10^{th}$ IEEE International Conference on Peer-to-Peer
Computing (IEEE P2P'10) in Delft, Netherlands, on August 25-27, 2010.}
}
\IEEEaftertitletext{\vspace{-0.33in}}
\maketitle

\begin{abstract}
Network latency and packet loss are considered to be an
important requirement for realistic evaluation of Peer-to-Peer
protocols. Dedicated clusters, such as Grid'5000, do not provide
the variety of network latency and packet loss rates that can be found in
the Internet. However, compared to the experiments performed on
testbeds such as PlanetLab, the experiments performed on dedicated
clusters are reproducible, as the computational resources are not
shared. In this paper, we perform experiments to study the impact of
network latency and packet loss on the time required to download a
file using BitTorrent. In our experiments, we observe a less than 15\%
increase on the time required to download a file when we increase the
round-trip time between any two peers, from 0~ms to 400~ms, and the
packet loss rate, from 0\% to 5\%. Our main conclusion is that the
underlying network latency and packet loss have a marginal impact on
the time required to download a file using BitTorrent. Hence,
dedicated clusters such as Grid'5000 can be safely used to perform
realistic and reproducible BitTorrent experiments.   
\end{abstract}

\begin{IEEEkeywords}
BitTorrent, Experiment, Performance, Clusters, Latency, Loss Rate.
\end{IEEEkeywords}

\section{Introduction}

A rich diversity in network latency and packet loss rates have become
essential for experimental evaluation of BitTorrent and other
communication protocols used in the Internet. The need for such
a diversity in network latency and packet loss rates is because of the
heterogeneous nature of the Internet~\cite{Floyd_2001_DiffSimInternet,
  Floyd_2008_ToolsTestbeds, Choffnes_2010_PitfallTestbed}. The
heterogeneity of the Internet is the primary motivation for the
creation of testbeds such as PlanetLab~\cite{PLANETLAB,
  Dischinger_2008_SatelliteLab}. However, due to the shared nature of
the PlanetLab platform, the results of the experiments performed on
PlanetLab are not reproducible~\cite{Spring_2006_PlanetLabMyths}. In
contrast, dedicated clusters, such as Grid'5000~\cite{Grid5000}, not
only offer a reproducible environment but also enable the scaling of
BitTorrent experiments by supporting a large number of BitTorrent
instances on a single machine. The primary shortcoming of experiments
performed on clusters is the absence of the diverse network latency
and packet loss rates that can be found in the Internet. As \emph{the
  impact of the network latency and packet loss on BitTorrent
  performance is not known}, there exists \emph{a dilemma while
  selecting a testbed for the BitTorrent experiments}.

The BitTorrent protocol uses TCP to efficiently distribute the
\emph{pieces} of a file to a large number of peers using peer-to-peer
(P2P) connections~\cite{Cohen_2008_BitTorrentProtocol}. The peers
contribute to the file distribution by uploading the pieces that have
been downloaded. During the file download, the peers exchange control
messages to select the pieces of the file to upload and the peers to
whom these pieces are to be uploaded. BitTorrent also allows the users
to limit the rate at which data is uploaded and downloaded. These rate
limits allow the users to restrict the network bandwidth that
BitTorrent can compete for during a BitTorrent session. As the users
can control the upload process by limiting the upload rates, and
BitTorrent uses control messages to decide the connections to upload
to, BitTorrent is inherently different from other TCP based file
transfer protocols such as HTTP and FTP.

Over the years BitTorrent performance has received considerable
attention~\cite{Qui_2004_ModelingP2P,Bharambe_2006_AIBNP,Chiu_2008_MinFDT}.
As these studies do not evaluate the interaction between TCP and
BitTorrent, the impact of network latency and packet loss rates
on BitTorrent performance is not known. Network latency and packet
loss introduce a \emph{ramp-up} period which is required by TCP to
attain the maximum upload rate that can be
achieved~\cite{RFC5681_2009_TCPCC, Mathis_1997_MBTCP}. BitTorrent
users that limit the upload rate can therefore limit the impact of
TCP ramp-up. Apart from the TCP throughput, the BitTorrent performance is
also dependent on the control messages exchanged by the peers. The
control messages generated by a peer can generate a delayed
response at a remote peer because of the network latency and the packet
losses. The delayed response can affect the various algorithms used by
BitTorrent, such as the peer selection algorithm which is used to
decide the peer to upload to. Due to the above reasons, the analytical
models for TCP cannot be directly used to provide the impact of network
latency and packet loss on BitTorrent performance.

In this paper, we use the download completion time, i.e, the time
required to download a file using BitTorrent, as a metric to study the
impact of network latency and packet loss. In our experiments, we
observe that network latency and packet loss have a marginal impact
(less than 15\%) on the download completion time of a file. We first
study the impact of network latency without packet loss. Network
latency causes not only delays in receiving control messages but also
TCP ramp-up. We therefore study the impact of network latency by
studying the impact of the delays in receiving control messages and
TCP ramp-up on the download completion time. We observe that the
download completion time, when the round-trip time (RTT) between
\emph{any two peers} in the torrent is 1000~ms, is not more than 15\%
of the download completion time when the RTT is 0~ms.

We then study the impact of network latency with packet loss. We observe
that an RTT of 400~ms \emph{between any two peers} and a packet loss
rate of 5\% does not increase the download completion time by more
than 15\% of the download completion time observed when the RTT is
0~ms and the packet loss rate is 0\%. We also observe that an RTT of
1000~ms \emph{between any two peers} with a packet loss rate of
5\% can increase the download completion time by more than 15\% of the
download completion time observed when the RTT is 0~ms and the packet
loss rate is 0\%. As an RTT greater than 400~ms between \emph{any two
  peers} is unrealistic, our results show that for upload rates
typically seen in the Internet, the download completion time is not
sensitive to the network conditions that can be found in the Internet;
\emph{dedicated clusters, such as Grid'5000, can be used to perform
  BitTorrent experiments}.

The remainder of this paper is structured as follows. The network
topologies and the technique used to emulate the latency and
packet loss are presented in \refSec{sec:Methodology}. We first
present the impact of latency without packet loss on download
completion time. In \refSec{sec:HomogeneousLatency}, we emulate the
same latency between any pair of peers in the torrent. We use this
topology to study the impact of TCP ramp-up and the impact of the
delays in receiving the BitTorrent control messages on the download
completion time. In \refSec{sec:HeterogeneousLatency}, we emulate
torrents to study the impact of network latency when the condition of
same network latency between any two peers is relaxed. We then study the
impact of network latency and packet loss in
\refSec{sec:ImpactLosses}. We finally conclude in
\refSec{sec:Conclusion}.

\section{Methodology}
\label{sec:Methodology}

In this paper, we use the terminology used by the BitTorrent
community. A \emph{torrent}, also known as a BitTorrent session or a
swarm, consists of a set of peers that are interested in having a
copy of the given file. A peer in a torrent can be in two states:
the \emph{leecher} state when it is downloading the file, and the
\emph{seed} state when it has a copy of the file being
distributed. The peers distribute the file in chunks called
\emph{pieces}; a piece is further split into \emph{blocks} to
facilitate the piece upload and download. A peer is said to
\emph{unchoke} a remote peer if it is uploading the blocks of
a piece. A \emph{tracker} is a server that keeps track of the peers
present in the torrent. For our experiments, we use a private torrent
with one tracker, one initial seed (henceforth called the seed), and
300 leechers. We use the \emph{download completion time}, the time
required by the leechers to download the file distributed using
BitTorrent, as the metric to study the impact of network latency and
packet loss. 

All the experiments were performed using an instrumented
version of the \emph{BitTorrent mainline client}~\cite{BTInstru} on
machines running Linux as the host operating system. The BitTorrent
mainline client internally uses TCP. We used TCP
Cubic~\cite{Ha_2008_Cubic}, the default TCP implementation for the
current series of the Linux kernel, for our experiments. The latest
version of uTorrent, a BitTorrent client, is based on uTP which uses
UDP as the transport layer protocol~\cite{Norberg_2010_uTP,
  LEDBAT}. As in the case of TCP, uTP has a window based congestion
control mechanism. The design of uTP also ensures that uTP ramp-up is
not faster than TCP~\cite{Norberg_2010_uTP}. The control messages 
generated by uTorrent using uTP are similar to those that are present
in BitTorrent clients that use TCP. Hence, we believe that the results
presented in this paper are valid for BitTorrent clients using uTP.     

In this section, we first present the various scenarios used in the
experiments. We then present the procedure to emulate the network
latency and packet loss. This is followed by an
overview of Grid'5000, the experimental platform we used. Finally, we
present the parameters used while performing the experiments.

\subsection{Experiment Scenarios}
\label{sec:ExpScen}

We now enumerate the scenarios used to study the impact of network
latency and packet loss on the download completion time of a file.
\begin{enumerate}
\item \emph{Scenario of Homogeneous Latency.} In this scenario, we
  emulate \emph{a fixed network latency between any two peers} in the
  torrent. The fixed network latency, though unrealistic, provides a
  controlled environment to study the impact of network latency on the
  download completion time of a file. We use this scenario to get an
  insight on the threshold of network latency beyond which the network
  latency affects the download completion time. The scenario also
  gives the download completion time when the maximum latency between
  any two peers in a torrent in known. The results of this study can also
  be used to give the impact of network latency when the hosts are
  geographically distributed and are connected with links that have
  negligible packet loss.  
\item \emph{Scenario of Heterogeneous Latency.} In this scenario we
  relax the condition of fixed network latency to emulate a more realistic
  network topology. We use this scenario to confirm that the results
  obtained in the scenario of homogeneous latency are valid even when
  the condition of fixed network latency between any two peers is relaxed. For
  this scenario, we group peers that have the same network latency
  among themselves in an emulated Autonomous System (AS). We assume
  these ASes to be fully meshed and that the inter-AS latency is
  greater than the intra-AS latency; we assume the network latency between the
  peers in a given AS is the same.  
\end{enumerate}

BitTorrent allows its users to limit the upload and download
rate. These rate limits restrict the network bandwidth that BitTorrent
can compete for with other applications such as Web browsers. In
this paper, we do not place any restrictions on the download rate. We
set the upload rate limit of the peers from a wide range of values,
from 10~kB/s to 100~kB/s. We use this range of upload rates as
Choffnes~\etal~\cite{Choffnes_2010_PitfallTestbed} show that 90\% of
the hosts present in public torrents upload at rates that are smaller
than 100~kB/s. In our experiments, we assume the same upload rate 
limit at all the leechers while studying the impact of network latency
and packet loss. We use the following torrent configurations to set
the limit on the upload rate of the peers.   
\begin{enumerate}
\item \emph{Slow Seed and Slow Leechers.}  In this scenario, we assume all
  the peers in the torrent have a low upload rate. We also assume the
  same upload rate at the seed and leechers. We performed two
  experiments for this scenario; we limit the upload to 10~kB/s for
  the first experiment and 20~kB/s for the second experiment.
\item \emph{Fast Seed and Slow Leechers.} Some torrents have seeds
  that upload faster than the leechers. Public torrents are also known
  to have leechers that are capable of high upload and download
  rates. In torrents where the seed favors fast
  leechers~\cite{Legout_2007_IncentiveBt}, these leechers are able to
  download the file faster than their slower peers. As these leechers
  are capable of downloading pieces faster than their slower peers,
  they act like a fast seed to the slow peers in their peer-set. We
  emulate torrents that have a fast seed by limiting the upload rate
  of the seed to 50~kB/s and the upload rate of the leechers to
  20~kB/s.
\item \emph{Fast Seed and Fast Leechers.} We perform these experiments to
  emulate torrents where all the peers are capable of high upload
  rates. We performed two experiments for this scenario; we limit the
  upload to 50~kB/s for the first experiment and 100~kB/s for the
  second experiment.
\end{enumerate}

\subsection{Emulation of Network Latency and Packet Loss}
\label{sec:EmulLatency}

\begin{figure}
\begin{centering}
\subfloat[][Network latency observed in the
Internet.]{\label{fig:LatencyInternet}\includegraphics[width=0.45\columnwidth]{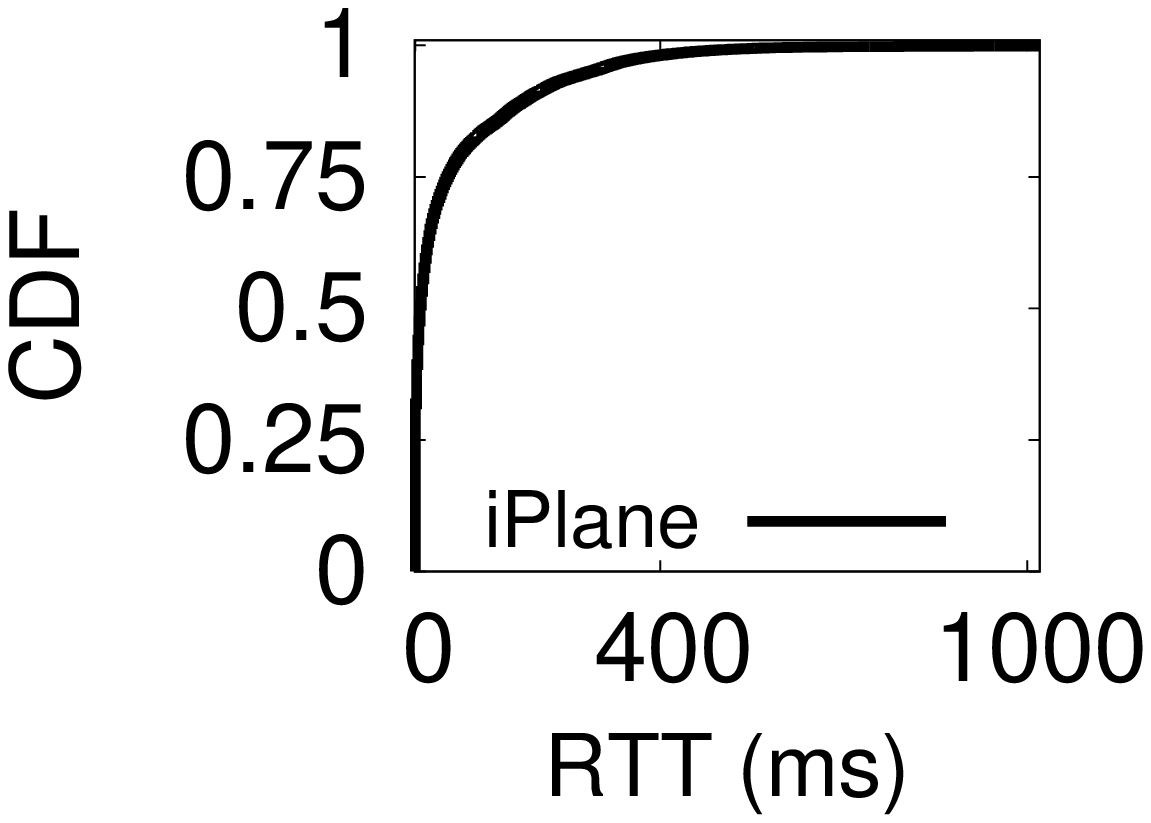}} 
\hspace{0.02\columnwidth}
\subfloat[][Loss rate observed in the Internet.]{\label{fig:LossRateInternet}\includegraphics[width=0.45\columnwidth]{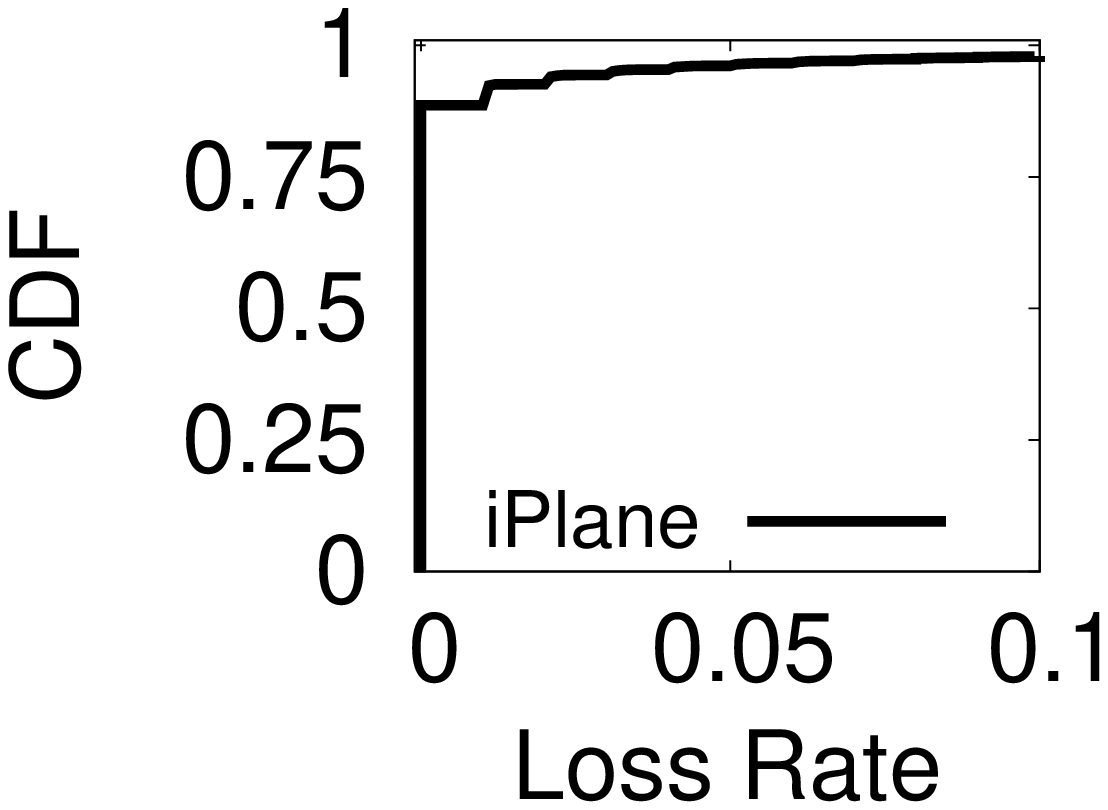}}
\caption{iPlane measurements of network latency and loss
  rate observed in the Internet. \scap{98\% of the links have an RTT
    less than 400~ms and 99.8\% of the links have an RTT less than
    1000~ms. 95\% of the links have a loss rate less than 0.05, i.e.,
    5\%.}}  
\label{fig:iPlaneMeasurements}
\end{centering}
\end{figure}

We use the Network Emulation (NetEm) module for the Linux
kernel~\cite{Hemminger_2005_NETEM} to emulate the network latency and
packet loss between the peers.  The NetEm module operates between the
TCP/IP implementation and the device driver for the network device. In
our experiments, the peers used the ethernet and the loopback device
to communicate with each other. The NetEm module emulates network
latency by en-queuing the packets at the ingress interface and the
egress interface of a network device; packets losses are emulated by
dropping the packets at the ingress and egress interfaces. When the
loss is on the egress interface, due to cross-layer optimizations
performed by the TCP implementation in Linux, TCP retransmits the
packet without considering it as a loss in the network. To avoid such
retransmissions, we introduce packet losses on the ingress interface
of the network device.

A shortcoming of NetEm is that the network latency and packet losses
are \emph{inherently} bound to the network device. NetEm can be used
to emulate network latency and losses for each connection, however such
topologies are hard to verify and manage. In this paper we set
the network latency and packet loss rates for a particular
network device. The network latency between a pair of peers is the sum
of the network latency at the devices used; the loss rate seen by a
packet is the loss rate at the ingress interface of the destination.

We use the publicly available iPlane
measurements~\cite{iPlane,Madhyastha_2006_IPlane} to obtain the values
of the network latency and loss rate to
emulate. \refFig{fig:iPlaneMeasurements} shows the distribution of
RTT and loss rate observed in 5 iPlane samples; each sample contains
the measurement of about $6*10^5$ links. In
\refFig{fig:LatencyInternet}, we observe that 98\% of the links have an
RTT less than 400~ms, and 99.8\% of the links have an RTT less than
1000~ms. We use these results to emulate a wide range of RTT values,
from 0~ms to 1000~ms, between a pair of peers in the
torrent. In \refFig{fig:LossRateInternet}, we observe
that 95\% of the links probed have a loss rate less than 0.05, i.e,
5\% packet loss. We emulate a 5\% packet loss on the ingress of each 
interface of the machines while studying the impact of packet
losses.

\subsection{Experiment Setup}
\label{sec:ExpParam}

We performed our experiments on the Grid'5000 experimental
testbed~\cite{Grid5000}. Grid'5000 consists of a grid of clusters
that are geographically distributed across France. Each cluster
consists of several machines connected by a dedicated and
high speed LAN; these clusters are inter-connected by high speed
links. We performed our experiments on \emph{a single cluster of
  Grid'5000}. In this cluster, we observe an RTT of less than 1~ms
between a pair of machines when network latency is not emulated. Due
to the very small network latency and the absence of packet loss 
between the machines in the LAN, the cluster does not reflect the
network conditions present in the Internet. Unlike other testbeds such
as PlanetLab, the machines in the cluster are not shared. Grid'5000
thus provides a reliable and robust platform for performing
reproducible experiments.     

\begin{figure}
\begin{centering}
\includegraphics[width=\columnwidth]{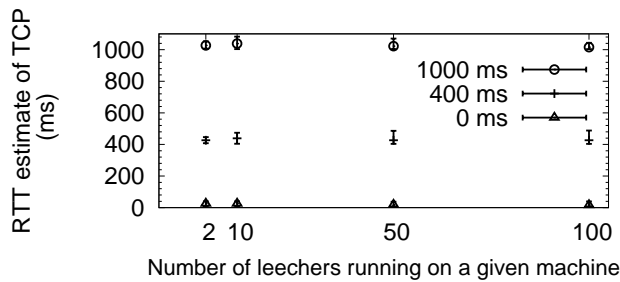}
\caption{Impact of the number of leechers running on a machine on the RTT
  estimate of TCP. \scap{The machines used can be used to run up to
    100 leechers.}}
\label{fig:ImpNumNodes}
\end{centering}
\vspace{-0.1in}
\end{figure}

We scale our experiments by running 100 leechers on a single
machine of the cluster. We performed the following test to ensure that
the machines can support up to 100 leechers while emulating the
desired network latency. We distribute a 50~MB file in a private torrent with
a single seed and a single tracker. The seed and tracker ran on one of
the machines of the cluster. We used another machine in the cluster to
run the leechers. We limit the upload rate of each of the peers to
100~kB/s and we vary the number of leechers running on a machine 
from 2 to 100. To study the impact of the number of leechers, we monitor 
the RTT estimate of TCP in the following manner. All the peers in the
torrent use the socket interface of TCP to communicate with each
other. The \verb,send, method of this interface is used by the peers
to send data to other peers in the torrent. On each call of
\verb,send,, we sample the RTT estimate of TCP by using the
\verb,TCP_INFO, option of the \verb,getsockopt,
method. \refFig{fig:ImpNumNodes} shows the average RTT estimate of TCP  
when we emulate an RTT of 0~ms, 400~ms, and 1000~ms, between the peers;
the error bars represent the minimum and maximum RTT estimate observed
in five iterations. We observe that the number of leechers running on
a given machine has a less than 15\% impact on the average RTT
estimate of TCP.   

The NetEm module buffers the TCP frames which are in flight for a time
period equal to the RTT being emulated. An RTT of 1 second
(1000~ms) for an upload rate of 100~kB/s would require 100~frames of
1~kB to be in flight. For our experiments, we use a buffer size of
$10^5$ frames to support 1000 frames of each of the 100 peers to be in
flight.

\begin{figure}
\begin{centering}
\includegraphics[width=0.85\columnwidth]{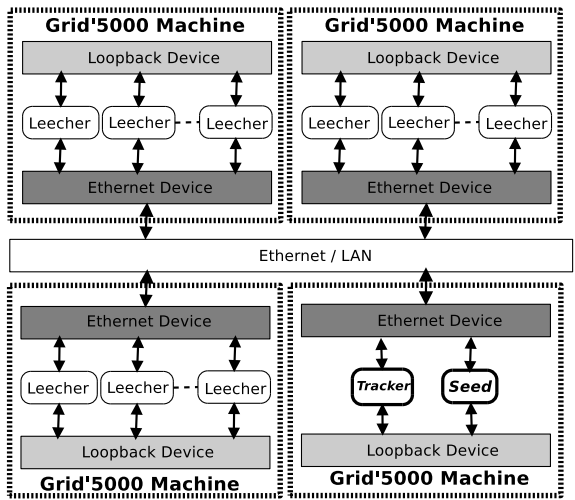}
\caption{Experiment setup. \scap{One machine for the tracker and the
    initial seed, and three machines each with 100 leechers.}}
\label{fig:Grid5KTopology}
\end{centering}
\end{figure}

We perform our experiments in a private torrent with one
tracker, one initial seed (henceforth called the seed), and a
flash crowd of 300 leechers. We distribute a 50 MB file in this
torrent. For our experiments, we assume that the peers remain in the
torrent until all the leechers have finished downloading the file. As
shown in~\refFig{fig:Grid5KTopology}, we use one machine for  
the tracker and the seed. We use three machines for the 300 leechers;
each machine runs 100 instances of the leechers. A pair of peers in the
torrent use either the loopback interface or the ethernet interface to
communicate with each other. 

\subsection{Impact of TCP Segmentation Offloading}
\label{sec:ImpactTSO}

The Maximum Segment Size (MSS) of TCP specifies the maximum payload
length that can be exchanged over a
connection~\cite{RFC5681_2009_TCPCC}. One factor that contributes to
the MSS is the Maximum Transmission Unit which is typically 1500 bytes
for devices used in LANs. The MTU is set to 16436 bytes for the
loopback interface on many Linux distributions. To ensure that the
payload length exchanged by the peers does not depend the interface
used by the connection, we set the MTU on the loopback interface to
1500 bytes. Despite this limit, we observed that a significant number
of TCP segments have a size larger than the MSS negotiated during
connection establishment. We observe these large segments because of
TCP Segmentation Offloading (TSO) which is enabled by default in the
2.6 series (the current series) of the Linux kernel. 

TCP Segmentation Offloading (TSO) enables the host machine to offload
some of the TCP/IP implementation, such as segmentation, and
calculation of IP checksum, to the network device. TSO also supports
the exchange of data in frames of sizes that can be greater that the
underlying MTU size~\cite{Mogul_2003_TCPOFFLOAD}. The increase in the
frame size can result in significant improvement in throughput; the
improvement depends on various factors such as CPU processing power
and the amount of data being
transferred~\cite{Freimuth_2005_SERVERTSO}. Clusters, such as  those
present in Grid'5000, include hosts that support TSO. 

\begin{figure}
\begin{centering}
\subfloat[][Payload length with TSO
Enabled. Increasing the RTT causes an increase in
payload length.]{\label{fig:PktLenTSOEnabled}\includegraphics[width=\columnwidth]{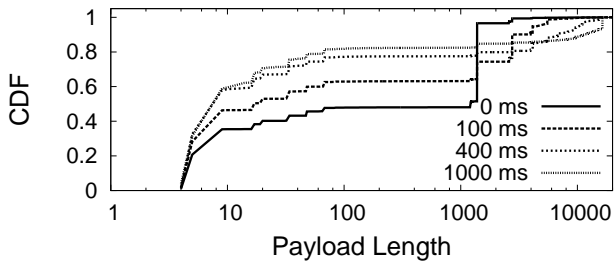}}\\
\vspace{-0.15in}
\subfloat[][Payload lengths observed in the WIDE
backbone. Maximum payload length less than 1500 bytes as TSO requires
hardware support on all the links on a connection.]{\label{fig:PktLenInternet}\includegraphics[width=\columnwidth]{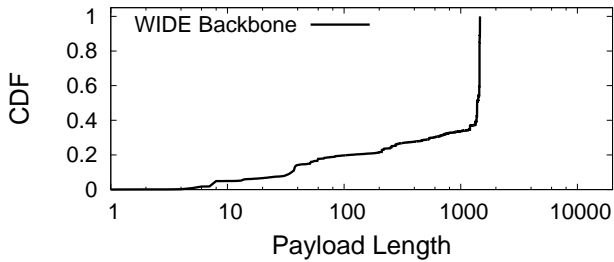}}\\
\vspace{-0.15in}
\subfloat[][Payload length with TSO Disabled and MTU set to 1500 bytes.]{\label{fig:PktLenTSODisabled}\includegraphics[width=\columnwidth]{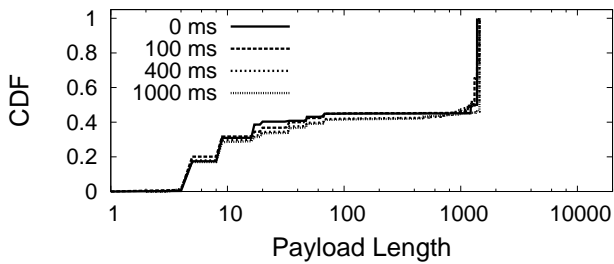}}
\caption{Impact of TSO on TCP payload length. \scap{Setting the MTU to 1500
  bytes and disabling TSO ensures that the maximum payload length is
  similar to that observed in the Internet.}}
\label{fig:ImpactTSO}
\end{centering}
\vspace{-0.2in}
\end{figure}

We now study the impact of RTT on the TCP payload length when TSO is
enabled. We use four machines to create a private torrent with 300
leechers, one tracker, and one initial seed, to distribute a 50~MB
file. We assume the same RTT between any two peers for the results
presented in \refFig{fig:PktLenTSOEnabled} and
\refFig{fig:PktLenTSODisabled}. We also limit the upload rate to
20~kB/s. \refFig{fig:PktLenTSOEnabled} shows the distribution of 
payload length for different RTT values when TSO is enabled; the
results present the distribution of the payload lengths observed in 5
iterations. We observe that an increase in the network latency between
the peers results in an increase in the number of TCP segments with a
large payload length. This was observed when the MTU was set to 1500
bytes indicating that TSO can result in payload lengths greater than
the MTU value specified. \refFig{fig:PktLenInternet} shows the payload
lengths from  the publicly available traces of the Internet traffic
observed in the WIDE backbone~\cite{MAWI}; the values presented are
from the sample taken on the WIDE backbone on November 29, 2009. In
\refFig{fig:PktLenTSOEnabled} and \refFig{fig:PktLenInternet}, we
observe that the maximum payload length of the packets sent over the
Internet is smaller than the maximum payload length observed in
Grid'5000 when TSO is enabled. We observe small payload lengths in
\refFig{fig:PktLenInternet} because hardware support on all the
intermediate devices is essential for exchange of large segments. When
TSO is disabled and the MTU is set to 1500 bytes,
\refFig{fig:PktLenTSODisabled} shows the distribution of the payload 
length for different RTT values; the results present the
distribution of the payload lengths observed in 5 iterations. We
observe that when TSO is disabled and the MTU is set to 1500 bytes,
the maximum payload length is similar to that observed in the
Internet. 

We set the MTU to 1500 bytes on the loopback interfaces and
disabled TSO on the ethernet and loopback interface for the
experiments presented in the \refSec{sec:HomogeneousLatency} and
\refSec{sec:HeterogeneousLatency}.  The outcome of the experiments
with TSO enabled are discussed with the impact of packet  loss in
\refSec{sec:ImpactLosses}. 

\section{Homogeneous Latency}
\label{sec:HomogeneousLatency}

\begin{figure}
\begin{centering}
\includegraphics[width=\columnwidth]{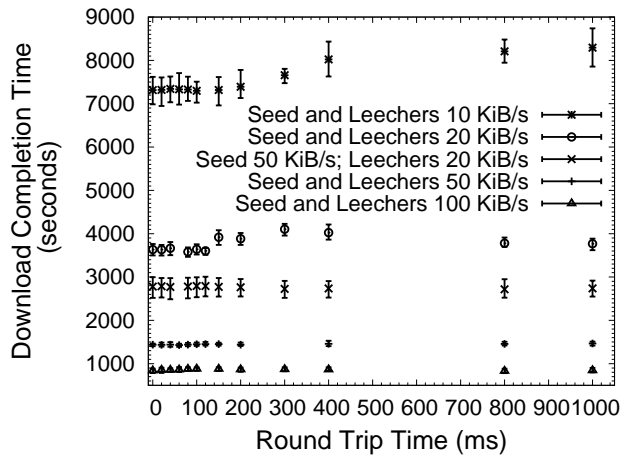}
\caption{Impact of latency on the average download completion
  time. The error bars indicate the minimum and maximum download
  completion time. \scap{The latency increases the download completion
    time by at most $15\%$.}}
\label{fig:SymmetricDownloadCompletion}
\vspace{-0.15in}
\end{centering}
\end{figure}

In this section, we assume the same network latency between any two
peers in the torrent because \emph{this scenario gives the worst case
  impact of a given network latency for a given upload rate
  limit}. The same network latency results in the same 
RTT between any two peers in the torrent because the sum of the other
delays, such as queuing delays at the intermediate routers, is less
than 1~ms in a Grid'5000 cluster. In each experiment, we choose an
RTT to emulate from a wide range of values, from 0~ms to 1000~ms. In
\refSec{sec:EmulLatency}, we observed that 99.8\% of the Internet
links have an RTT less than 1000~ms. \emph{An RTT of 1000~ms between
any two peers thus gives a worst case impact of the network latency
that can be observed in the Internet}. We use a wide range of values,
from 10~kB/s to 100~kB/s, to set the upload rate limit of the peers in
the torrent; we assume the same upload rate limit for all the leechers
in the torrent. For a given upload rate limit,
\refFig{fig:SymmetricDownloadCompletion}  
shows the impact of the RTT between any two peers on the download
completion time. Each point on the plot is the average download
completion time in 10 iterations; the error bars indicate the minimum
and maximum download completion time observed in 10 iterations. The
two factors that can be affected by network latency and cause an
increase in the download completion time are TCP ramp-up and the
delays in receiving the control messages. We study the impact of
network latency on the download completion time by studying the impact
of network latency on these two factors in \refSec{sec:ImpactRampUp}
and \refSec{sec:ControlMessagesImpact} respectively.

In \refFig{fig:SymmetricDownloadCompletion}, for a given upload rate
limit from 10~kB/s to 100~kB/s, we observe that an RTT of up to
1000~ms between any two peers does not increase the average download
completion time by more than 15\% of the average download completion
time when the RTT is 0~ms. When the maximum upload rate of the seed
and leechers is limited to 10~kB/s,
\refFig{fig:SymmetricDownloadCompletion} shows that the download
completion time increases when the RTT is greater than 200~ms. In
\refSec{sec:ImpactRampUp}, we show how TCP ramp-up is responsible for
this increase in the download completion time. When the upload rate of
all the peers is limited to 20~kB/s, in
\refFig{fig:SymmetricDownloadCompletion}, we observe that the download
completion time is not a monotonous function of the RTT. Peers having
an RTT of 1000~ms have a lower download completion time compared to
peers having an RTT of 400~ms. We show that this is the impact of
network latency on the delays in receiving BitTorrent control messages
in \refSec{sec:ControlMessagesImpact}.

An RTT of 1000~ms between any two peers gives the worst case impact
of network latency. \refFig{fig:SymmetricDownloadCompletion} thus
shows that \emph{network latency has a marginal impact, less than
  15\%, on the average download completion time for the upload rates
  observed in public torrents}.

\subsection{Impact of TCP Ramp-Up}
\label{sec:ImpactRampUp}

\begin{figure}
\begin{centering}
\subfloat[][Distribution of the time between successive send
system calls at the \emph{leechers} when RTT between peers is 0~ms.]{\label{fig:Leecher0}\includegraphics[width=\columnwidth]{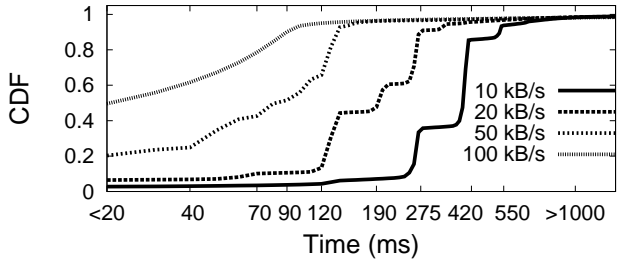}}
\hspace{0.05\columnwidth}
\subfloat[][Distribution of the time between successive send
system calls at the \emph{leechers} when RTT between peers is 400~ms.]{\label{fig:Leecher400}\includegraphics[width=\columnwidth]{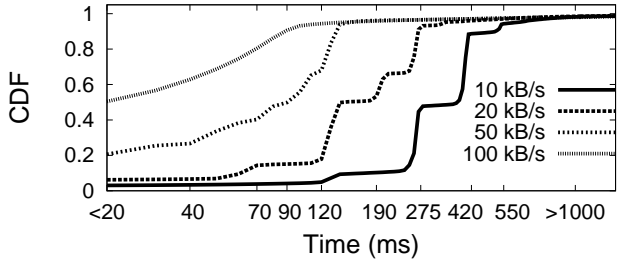}}
\caption{Distribution of the time between successive send system calls
  at the leechers (semi-log scale). \scap{The RTT does not affect the upload
  process at the seed and the leechers for high upload rates.}}
\label{fig:IATLeecher}
\end{centering}
\vspace{-0.25in}
\end{figure}

The BitTorrent application uses the \verb,send, system call of the
socket interface of TCP to upload the pieces of a file. BitTorrent
limits the upload rate for a given TCP connection by periodically
calling the \verb,send, system call. The time between successive calls
of \verb,send, is high when the upload rate is low. Each call of
\verb,send, typically results in creation of a TCP
segment which is transmitted. If the acknowledgment for the
transmitted segment arrives before the subsequent call of \verb,send,,
then the upload rate of the application does not require a congestion
window ramp-up. \emph{A ramp-up is required on the TCP connections
  where the time between successive calls of} \verb,send, \emph{is
  smaller than the RTT}. BitTorrent also enables a peer to simultaneously
upload pieces of the file to many peers of a torrent in parallel. An
increase in the number of parallel unchokes results in an increase in
the time between successive calls of \verb,send,. This is because the
upload rate limit is divided over the connections used for the
unchoke.

We present the distribution for the time between successive calls of
the \verb,send, at all the leechers (over 5 iterations) in
\refFig{fig:IATLeecher}; the RTT between any two peers is 0~ms in
\refFig{fig:Leecher0} and 400~ms in \refFig{fig:Leecher400}.
For an upload rate of 10~kB/s, in \refFig{fig:Leecher0} and
\refFig{fig:Leecher400}, we observe that the time between
successive \verb,send, calls is greater than 200~ms for a significant
number of calls. For these calls, when the RTT between the peers is
less than or equal to 200~ms a ramp-up shall not be required. Due to
the absence of ramp-up, in \refFig{fig:SymmetricDownloadCompletion}
for an upload rate limit of 10~kB/s, we observe that an RTT less than
or equal to 200~ms does not increase the average download completion
time by more than 5\% of the average download completion time observed
when the RTT is 0~ms. For the upload rate limit of 10~kB/s, an
RTT larger than 200~ms may require a TCP ramp-up. Similarly, for the
upload rate limits of 20~kB/s, 50~kB/s, and 100~kB/s, in
\refFig{fig:IATLeecher}, we observe that ramp-up may be   
required for an RTT greater than 120~ms, 20~ms, and 20~ms
respectively. Hence, for an upload rate limit in the range of 10~kB/s
to 100~kB/s, TCP ramp-up will be required for an RTT greater than
200~ms. In \refFig{fig:SymmetricDownloadCompletion} for an RTT 
of up to 1000~ms, we observe that TCP ramp-up does not increase the
average download completion time by more than 15\% of the download
completion time observed the RTT is 0~ms. Thus, \emph{TCP ramp-up has
  a marginal impact, less than 15\%, on the average download
  completion time for the upload rates observed in public torrents}.  

\subsection{Impact of Latency on Control Messages}
\label{sec:ControlMessagesImpact}

In \refFig{fig:SymmetricDownloadCompletion} for an upload rate of
20~kB/s, we observe that the download completion time is not a
monotonous function of the RTT. The download completion time when the
RTT is 1000~ms is smaller than the download completion time when the
RTT is 400~ms. We now present the reasons for this behavior. We would
like to comment that this discussion is specific to the implementation
of the BitTorrent client we used in the experiments and that
\emph{these observations may not be true for other BitTorrent
  clients that are available}.

Once every 10 seconds, a seed selects from its peer-set a set of
leechers to unchoke. The leechers are selected based on their download
rate and time since the start of their last unchoke. During an
unchoke, the leecher downloading the piece requests for multiple
blocks of a piece to pipeline the blocks for the unchoke. The number
of blocks requested is a function of the estimated download rate at
the leecher. This estimate of the download rate is a moving average
and it grows slowly to attain the rate at which the seed uploads to
the given leecher. This growth is even slower if the connection
requires a TCP ramp-up. The slow growth in the estimated download rate
thus results in a slow increase in the number of blocks that are in
the pipeline at the seed. We observe that this slow increase, along
with the latency in receiving the block requests, results in time
periods during which the seed unchoking the leecher is idle, i.e.,
awaiting block requests. If the leecher selection algorithm of the
seed is invoked during these \emph{idle periods}, then the leecher
\emph{will not be selected} for an unchoke resulting in an abrupt
termination of the unchoke. We observe that this abrupt termination of
the unchoke occurs frequently when the upload rate limit is 20~kB/s
and when the RTT between the peers is greater than 120~ms and less
than 800~ms. 

\begin{figure}
\begin{centering}
\subfloat[][Number of pieces available with the leechers when the
upload rate is limited to 20~kB/s.]{\label{fig:Have20}\includegraphics[width=\columnwidth]{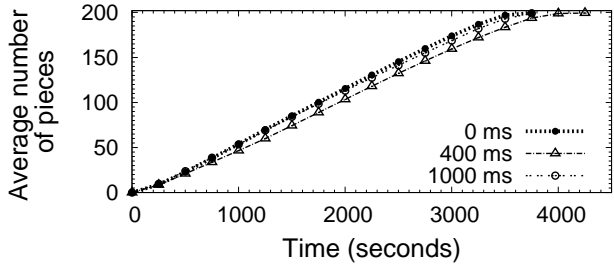}}
\hspace{0.05\columnwidth}
\subfloat[][Number of pieces available with the leechers when the
upload rate of the leechers is limited to 20~kB/s and that of the
seed is limited to 50~kB/s.]{\label{fig:Have5020}\includegraphics[width=\columnwidth]{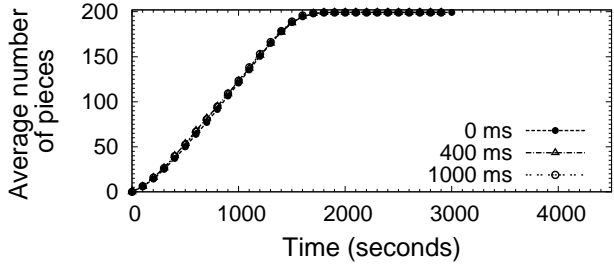}}
\caption{Evolution of pieces available at the leechers. \scap{Latency has
  marginal impact on the pieces available with the leechers.}}
\label{fig:RTTHaveImpact}
\end{centering}
\vspace{-0.15in}
\end{figure}

We now show how the abrupt termination of an unchoke affects the
availability of pieces at the leechers. Once a piece is downloaded,
the leecher can then upload this piece to other leechers in its
peer-set. After the piece download, the leecher sends a \emph{HAVE}
message with the piece identifier to the peers in its peer-set. The
\emph{HAVE} message indicates that the piece can be downloaded from
this leecher. In \refFig{fig:RTTHaveImpact}, we use the \emph{HAVE}
messages sent by the leechers to show the evolution of the average
number of pieces (over 10 iterations) that are available at the
leechers. For reasons mentioned above, when the upload rate limit of
the seed and leechers is 20~kB/s, an RTT of 400~ms between the seed
and a leecher typically results in the abrupt termination of the
unchoke. The abrupt termination results in an incomplete download of a
piece and affects the availability of pieces as shown in
\refFig{fig:Have20}. We do not observe abrupt termination of the
unchoke when the RTT is 0~ms and when the RTT is 1000~ms. Hence, the
average download completion time when the RTT is 400~ms is greater
than the average download completion observed when the RTT is
1000~ms. 

We do not observe abrupt termination of unchokes when we have
a fast seed (upload rate limited to 50~kB/s) in a torrent with slow
leechers (upload rate limited to 20~kB/s); \refFig{fig:Have5020} shows
the evolution of pieces for this torrent configuration. We observe a
similar evolution for the \emph{HAVE} messages when the upload rate at
the peers is limited to 50~kB/s and 100~kB/s.

\subsection{Summary}

In this section, we emulated a large range of RTT values, from 0~ms to
1000~ms, to study the impact of network latency on the download
completion time. We use an RTT of 1000~ms to give the worst case
impact of network latency for a given upload rate. We observe that an
RTT of up to 1000~ms between any two peers has a marginal impact, less
than 15\%, on the average download completion time of a file. For an
upload rate of 20~kB/s, we observe that the download completion time is
not a monotonous function of the RTT. This behavior emphasizes
\emph{that the models for TCP throughput cannot be directly used to
  study the impact of network latency on the time required to download
  a file using BitTorrent}. 

In the next section, we relax the condition of same latency between any
two peers in the torrent to confirm that the scenario of same latency
gives us the worst case impact of network latency.

\section{Heterogeneous Latency}
\label{sec:HeterogeneousLatency}

We now present results to confirm that the observations made in
\refSec{sec:HomogeneousLatency} are valid even when the condition of
fixed latency is relaxed. We relax the condition of same latency
between any two peers by emulating ASes in the following
manner. Public torrents have peers that are spread out
geographically. A pair of peers in the same AS typically have a
smaller network latency compared to a pair of peers that are present
in different ASes. For our experiments, we use a private torrent with
peers distributed in emulated ASes. For our experiments, we
emulate an AS using one machine. We assume the same intra-AS
latency and we also assume that the intra-AS latency is less than the 
inter-AS latency. We assume that all the ASes are fully meshed.

\subsection{Emulation of ASes}

\begin{table}
\begin{centering}
\begin{tabular}{|c|c|c|}
\hline
AS & Latency on  & Latency on \\
   & Loopback (ms) & Ethernet (ms)  \\
\hline
$AS_{1}$ & 2 & 5 \\
\hline
$AS_{2}$ & 5 & 15 \\
\hline
$AS_3$ & 10 & 25 \\
\hline
$AS_4$ & 25 & 100 \\
\hline
$AS_5$ & 50 & 100\\
\hline
\end{tabular}
\caption{Latency values on the loopback and ethernet device while
  emulating an AS on a machine.}
\label{tab:LatencyAS}
\end{centering}
\end{table}
\begin{table}
\begin{centering}
\begin{tabular}{|c|c|c|c|c|c|}
\hline
     & $AS_{1}$ & $AS_{2}$ & $AS_3$ & $AS_4$ & $AS_5$\\
\hline
$AS_{1}$ & 8 ms  & 40 ms  &   60 ms &  210 ms & 210 ms\\
\hline
$AS_{2}$ & 40 ms & 20 ms  &   80 ms &  230 ms & 230 ms\\
\hline
$AS_3$ & 60 ms  & 80 ms  &   40 ms &  250 ms & 250 ms \\
\hline
$AS_4$ & 210 ms & 230 ms &  250 ms&  100 ms & 400 ms\\
\hline
$AS_5$ & 210 ms & 230 ms &  250 ms &  400 ms & 200 ms\\
\hline
\end{tabular}
\caption{RTT between a pair of leechers. \scap{ RTT between a leecher in $AS_1$
  and a leecher in $AS_5$ is 210 ms.}}
\label{tab:RTTAS}
\end{centering}
\end{table}

As in the case of homogeneous latency, we use four machines in each
of the experiments. We use a private torrent with 300 leechers, one
tracker, and one initial seed. We emulate four ASes: three ASes each
with 100 leechers, and the fourth AS to emulate the AS of the seed and
the tracker. The four ASes used in these experiments were chosen from
a set of five ASes ($AS_{1}$, $AS_{2}$, $AS_{3}$, $AS_{4}$, and
$AS_{5}$).

We now present an explanation for emulating these five
ASes. In \refFig{fig:SymmetricDownloadCompletion},
for an upload rate of 20~kB/s, we observe that an RTT smaller than
120~ms between any two peers has a smaller impact on the download
completion time as compared to an RTT larger than 120~ms. In three of
the five ASes, namely $AS_{1}$, $AS_{2}$, and $AS_{3}$, the RTT
between a pair of peers in these three ASes is less than 120~ms. The
RTT between a pair of peers in $AS_{4}$ is less than 120~ms; the RTT
between a peer in $AS_{4}$ and any other peer is greater than
120~ms. Finally, a peer in $AS_{5}$ has an RTT greater than 120~ms
with any other peer. As we use one machine to emulate an AS, peers in 
the same AS use the loopback device to communicate with each other;
the peers use the ethernet device to communicate with all the other
peers in the torrent. We emulate the latency values given in
\refTab{tab:LatencyAS} for the ethernet and loopback device while
using a machine to emulate an AS.

We now give an example to show how \refTab{tab:LatencyAS} can be used
to find the RTT between a pair of leechers. The RTT between a leecher
in $AS_1$ and a leecher in $AS_2$ is 40~ms (5+15+15+5) as the leechers
use the ethernet device to communicate with each other. The RTT
between a pair of leechers in $AS_1$ is 8~ms (2+2+2+2) as the
leechers use the loopback device to communicate with each
other. \refTab{tab:RTTAS} gives the RTT values between all such pairs
of leechers.

\begin{table}
\begin{centering}
\begin{tabular}{|c|c|c|c|c|c|}
\hline
     & $AS_{1}$ & $AS_{2}$ & $AS_3$ & $AS_4$ & $AS_5$\\
\hline
$AS_{1}'$ & 20 ms  & 40 ms  &   60 ms &  210 ms & 210 ms\\
\hline
$AS_{2}'$ & 40 ms & 60 ms  &   80 ms &  230 ms & 230 ms\\
\hline
$AS_3'$ & 60 ms  & 80 ms  &   100 ms &  250 ms & 250 ms \\
\hline
$AS_4'$ & 210 ms & 230 ms &  250 ms&  400 ms & 400 ms\\
\hline
$AS_5'$ & 210 ms & 230 ms &  250 ms &  400 ms & 400 ms\\
\hline
\end{tabular}
\caption{RTT between the seed and the leechers in the torrent. \scap{$AS_{i}'$
  indicates that seed is placed in as AS with latency values similar
  to $AS_{i}$. RTT between the seed in $AS_1'$ and a peer in $AS_1$ is
  20 ms.}}
\label{tab:RTTASSeed}
\end{centering}
\end{table}

For our experiments, we assume that the seed and the tracker are
placed in a dedicated AS with no leechers. We use $AS_i'$ to denote
that the seed and the tracker are placed in an AS with the same
latency values as $AS_i$. For example, $AS_{1}'$ implies that the seed
and tracker are placed in an AS having the same latency values as
$AS_{1}$. \refTab{tab:RTTASSeed} gives the RTT values between the seed
and the leechers.

\subsection{Presentation and Discussion of Results}

\begin{figure}
\begin{centering}
\subfloat[][Download completion time for leechers present in $AS_{1}$, $AS_{2}$, and
$AS_{3}$. The difference in the average download completion time is
less than 15\%.]{\label{fig:Asym20AS123}\includegraphics[width=\columnwidth]{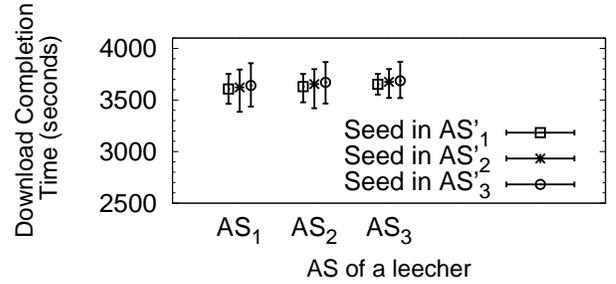}}

\subfloat[][Download completion time for leechers present in $AS_{1}$, $AS_{3}$, and
$AS_{4}$. An RTT of 400~ms between a leecher in $AS_4$ and the seed
in $AS_4'$ does not increase the average download completion time by
more than 15\%.]{\label{fig:Asym20AS134}\includegraphics[width=\columnwidth]{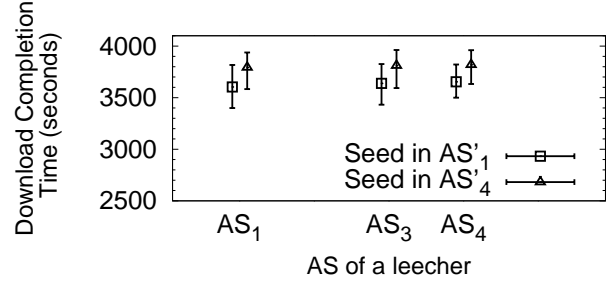}}

\subfloat[][Download completion time for leechers present in $AS_{1}$, $AS_{3}$, and
$AS_{5}$. An RTT of 400~ms between a leecher in $AS_4$ and the seed
in $AS_4'$ does not increase the average download completion time by
more than 15\%.]{\label{fig:Asym20AS135}\includegraphics[width=\columnwidth]{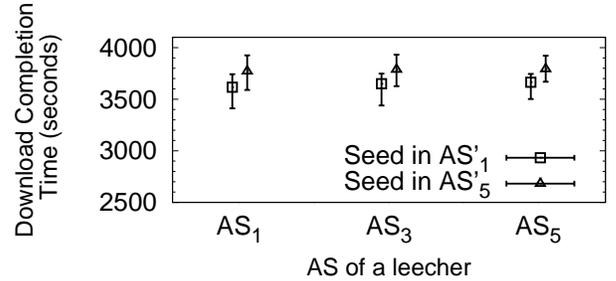}}
\caption{Download completion time of a 50 MB file by leechers in a
  given AS when the maximum upload rate of all the peers is
  20~kB/s. \scap{Despite the wide range of RTT values emulated, the
  difference in the download completion time is less than 15\%.}}
\label{fig:AsymUp20}
\end{centering}
\end{figure}

\begin{figure}
\begin{centering}
\includegraphics[width=\columnwidth]{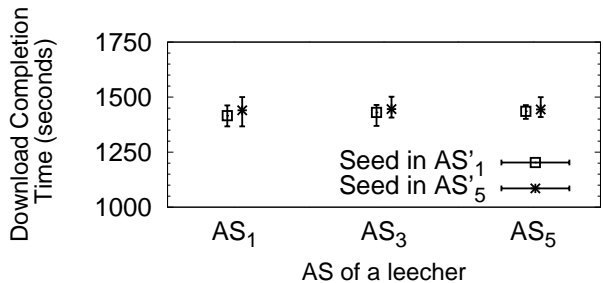}
\caption{Download completion time of a 50 MB file by leechers present
  in a given AS when the maximum upload rate of all the peers is
  50~kB/s. \scap{Despite the wide range of RTT values emulated, the
  difference in the download completion time is less than 15\%.}}
\label{fig:AsymUp50}
\end{centering}
\end{figure}

\refFig{fig:AsymUp20} show the impact of heterogeneous latency on the
download completion time of a 50 MB file when the upload rate of the
peers is limited to 20~kB/s. The impact of network latency when the
upload rate limit is 50~kB/s is presented in \refFig{fig:AsymUp50}. In
\refFig{fig:AsymUp20} and 
\refFig{fig:AsymUp50}, the X-axis represents the AS of the leechers
present in the torrent, and the Y-axis represents the download
completion time in seconds. The figures present the average download
completion time over 10 iterations; the error bars indicate the
minimum and maximum download completion time observed in 10
iterations.

\refFig{fig:Asym20AS123} shows the outcome of three experiments
with leechers in $AS_{1}$, $AS_{2}$, and $AS_{3}$. For a given
experiment, the seed was either in $AS_{1}'$, $AS_{2}'$, or
$AS_{3}'$. Despite the different RTT values between the peers, in the
three experiments presented in \refFig{fig:Asym20AS123}, we observe
that the difference in the average download completion time is less
than 15\%. According to \refTab{tab:RTTAS}, 
the RTT between any two peers in these experiments was less than
120~ms. For the experiments presented in \refFig{fig:Asym20AS134} and
\refFig{fig:Asym20AS135}, a peer in $AS_1$ or $AS_3$ and 
another peer in $AS_4$ or $AS_5$ have an RTT greater than 120~ms. 
Despite the wide range of RTT values, we observe that
the average download completion time in \refFig{fig:Asym20AS134} and
in \refFig{fig:Asym20AS135} is not more than $15\%$ of the average
download completion time observed in \refFig{fig:Asym20AS123}, i.e.,
when the RTT between a pair of peers is less than 120~ms. 

According to \refTab{tab:RTTAS}, a peer in $AS_5$ and the seed in
$AS_5'$ have an RTT of 400~ms. We observe that the average download
completion time in \refFig{fig:Asym20AS135} is smaller than the
average download completion time observed in
\refFig{fig:SymmetricDownloadCompletion} for an upload rate limit of
20~kB/s and an RTT of 400~ms. This shows that \emph{the scenario of
homogeneous latency can be used to give a worst case impact of network
latency for a given upload rate}. When the upload rate limit is set to
50~kB/s, in \refFig{fig:AsymUp50} we observe that the difference
in the average download completion time is less than 15\%.     

\subsection{Summary}

In this section, we relax the condition of fixed latency between any
two peers in a torrent. We observe that an RTT of up to 400~ms has a
marginal impact, less than 15\%, on the average download completion
time. These observations show that the \emph{upload process at the 
  peers is not sensitive to the variations in the TCP throughput due
  to the change in latency}. These observations also confirm that, for
a given upload rate among the peers, \emph{the scenario of homogeneous
latency provides an upper bound on the download completion time of a
file when the maximum latency between any two peers in a torrent is
known}.

\section{Impact of Packet Loss}
\label{sec:ImpactLosses}

We now present the impact of packet loss on the download completion
time of a file. For our experiments, we emulate a 5\% packet loss on
the ingress interface of the loopback and ethernet devices of the
machines.

In \refTab{tab:ImpactLossTSO}, we present the average download
completion time of a 50~MB file for a given upload rate limit and a
given RTT between any two peers. As in the case of homogeneous
latency, we consider a torrent consisting of a one tracker, one seed,
and a flash crowd of 300 leechers. We observe the download completion
time for the following network conditions.
\begin{enumerate}
\item \emph{Homogeneous latency and TSO is disabled}. We do not emulate
  packet losses in this scenario. As packet losses are not
  emulated, this setting gives the outcome of experiments performed on
  clusters that do not support TSO. The results of this scenario are
  discussed in detail in \refSec{sec:HomogeneousLatency}.
\item \emph{Homogeneous latency and TSO is enabled}. We do not emulate
  packet losses in this scenario. The results present the outcome of
  experiments performed without emulating packet loss on clusters that
  support TSO.
\item \emph{Homogeneous latency with a loss rate of 5\% and TSO
  disabled}. We use this scenario to emulate network conditions present
  in the Internet.
\end{enumerate}

\begin{table}
\begin{center}
\begin{tabular}{|c|c||c|c|c|}
\cline{3-5}
\multicolumn{2}{}{ }& \multicolumn{3}{|c|}{\bf Average Download Completion Time}\\
\hline
Upload& RTT & {TSO disabled} & { TSO enabled} & { TSO disabled}\\
Rate & & { Loss Rate 0\%} & { Loss Rate 0\%} & { Loss Rate 5\%}\\
\hline
10 &  0 ms & 7314.7 s & 7268.4 s & 7359.1 s \\
\cline{2-5}
kB/s & 400 ms & 8006.0 s & 7823.4 s & 8183.6 s\\
\cline{2-5}
 & 1000 ms & 8274.19 s  & 8060.6 s & {8827.6 s}\\
\hline
20 &  0 ms & 3634.9 s & 3728.3 s & 3711.2 s\\
\cline{2-5}
kB/s & 400 ms &4023.9 s & 3985.3 s & 4034.3 s \\
\cline{2-5}
 & 1000 ms & 3768.3 s & 3796.6 & 4102.7 s \\
\hline
50 &  0 ms & 1437.96 s& 1433.7 s & 1432.7s \\
\cline{2-5}
kB/s & 400 ms & 1457.2 s& 1463.9 s& 1476.9\\
\cline{2-5}
 & 1000 ms & 1466.9 s& 1470.5 s & 1638.2 s\\
\hline
100 &  0 ms & 838.9 s & 828.9 s & 832.7 s\\
\cline{2-5}
kB/s &  400 ms & 863.4 s & 860.4 s & 940.20 s\\
\cline{2-5}
 &  1000 ms & 844.5 s & 865.4 s & {1619.87 s} \\
\hline
\end{tabular}
\end{center}
\caption{Impact of RTT and loss rate on download completion
  time. \scap{An RTT of up to 400~ms and the loss rate of up to 5\%
    does not increase the average download completion time by more
    than 15\% of the average download completion time observed when
    the RTT is 0~ms and the loss rate is 0\%.}}
\label{tab:ImpactLossTSO}
\vspace{-0.15in}
\end{table}

In \refSec{sec:HomogeneousLatency}, we observed the impact of network
latency when TSO is disabled and packet loss is not emulated. We observe
that for an RTT of up to 1000~ms, TCP ramp-up and the delays in
receiving the control messages have a marginal impact, less than 15\%,
on the average download completion time of a file.

When TSO in enabled and packet losses are not emulated, we observe
that the impact of network latency on the download completion time is
similar to that observed in the case of homogeneous latency, which has
been presented in \refSec{sec:HomogeneousLatency}. These results show
that \emph{BitTorrent experiments performed on clusters that support TSO 
shall produce results that are similar to those performed on clusters
that do not support TSO}.

We now discuss the scenario of homogeneous latency with packet loss
when TSO is disabled. While emulating a packet loss, each packet
en-queued by NetEm is dropped with a probability controlled by the
loss rate. A packet loss result in retransmission of the packet by
the source. The source reduces the congestion window in response
to the loss and then ramps-up its congestion window to attain the
desired upload rate. We observe that for an upload rate limit of
20~kB/s and 50~kB/s, an RTT of 1000~ms and a loss rate of $5\%$
between any two peers does not increase the average download completion time
by more than $15\%$ of the average download completion time observed
when \emph{neither latency nor packet loss was emulated}. However,
when the upload rate limit is 10~kB/s or 100~kB/s, the RTT between any
two peers is 1000~ms, and when the loss rate is 5\%, we observe that the
average download completion time is more than 15\% of the download
completion time when the RTT is 0~ms and the loss rate is 0\%. An RTT
of 1000~ms between any two peers is unrealistic as 99.8\% links probed
by iPlane have an RTT less than 1000~ms. However, \emph{an RTT of up
to 400~ms is realistic} as 98\% of the links probed by iPlane have an
RTT less than 400~ms. For the upload rate limit from 10~kB/s to
100~kB/s, when the RTT between any two peers is 400~ms and the loss
rate is 5\%, we observe that the download completion time is not more
than 15\% of the download completion time observed when the RTT is
0~ms and loss rate is 0\%. In \refSec{sec:HeterogeneousLatency} we
observe that the scenario of homogeneous latency gives an upper bound
on the impact of a given network latency. Therefore, the results
presented in \refTab{tab:ImpactLossTSO} show that, for an upload rate
limit from 10~kB/s to 100~kB/s, \emph{a loss rate of up to 5\% and an
  RTT of up to 400~ms between any two peers in the torrent has a
  marginal impact, less than 15\%, on the average download completion
  time of a file.} 

\section{Concluding Remarks}
\label{sec:Conclusion}

In this paper we present the impact of network latency and packet
loss on the download completion time of a file distributed using
BitTorrent. We use the download completion time as the metric for
evaluation because the BitTorrent users are primarily interested in
the download completion time of a file.

We first studied the impact of network latency on the download
completion time. For a given upload rate limit, we emulated the same
latency among the peers to give a worst case impact of network latency
on the download completion time. The download completion time can be
affected by TCP ramp-up and the delays in receiving the BitTorrent
control messages. We therefore studied the impact of network latency
on the download completion time by studying the impact of network
latency on TCP ramp-up and the delays in receiving the control
messages. For our experiments, we varied the upload rate limit from
10~kB/s to 100~kB/s and the RTT from 0~ms to 1000~ms. We observe that
the TCP ramp-up and the delays in receiving the BitTorrent control
messages only have a marginal impact, less than 15\%, on the average
download completion time. The high RTT values used in our experiments
also emulate torrents with peers that are not only geographically
apart but also connected with high capacity links that support the
BitTorrent upload rate without causing congestion. Our results show
that \emph{experiments performed on Grid'5000 give results similar to
those performed on testbeds such as PlanetLab that have geographically
distributed hosts that are connected by high capacity links}. We also
study the impact of network latency on the delays in receiving
the control messages; this impact cannot be captured using the 
traditional models for TCP throughput.

We then studied the impact of packet loss by emulating a loss rate of
5\%. For the upload rates seen in public torrents, from 10~kB/s to
100~kB/s, we observe that realistic RTT values of up to 400~ms and a
packet loss rate up to 5\%, have a marginal impact, less than 15\%, on
the average download completion time.  We performed our experiments
over a wide range of RTT values, from 0~ms to 1000~ms, and a wide
range of packet loss rates, from 0\% to 5\%, to study the impact of
network latency and packet loss. Our results show that
\emph{experiments can be performed on dedicated clusters, such as
  those present in Grid'5000, without explicitly emulating latency and
  packet loss between the peers in a torrent}.    

We also studied the impact of using devices that support TSO on the
outcome of BitTorrent experiments performed on clusters. Our results
show that BitTorrent experiments performed on clusters that support TSO
produce results that are similar to those performed on dedicated
clusters that do not support TSO.  

Our main conclusion is that, for upload rates seen in public torrents,
network latency and packet loss have a marginal impact on the download
completion time of a file, hence, dedicated clusters such as Grid'5000
can be safely used to perform realistic and reproducible BitTorrent
experiments.   

\section{Acknowledgment}

Experiments presented in this paper were carried out using the Grid'5000
experimental testbed, being developed under the INRIA ALADDIN development
action with support from CNRS, RENATER and several Universities as well
as other funding bodies (see https://www.grid5000.fr).

\end{document}